\documentclass[twocolumn,twoside,preprintnumbers]{revtex4}%
\usepackage{amssymb}
\usepackage{amsmath}
\usepackage{epsfig}
\usepackage{graphicx}
\usepackage{amsfonts}%
\begin{document}
\title{Survival probability of surface excitations in a 2\textit{d} lattice:
non-Markovian effects and Survival Collapse.}
\author{E. Rufeil Fiori}
\email{rufeil@famaf.unc.edu.ar}
\affiliation{Facultad de Matem\'{a}tica, Astronom\'{\i}a y F\'{\i}sica, Universidad
Nacional de C\'{o}rdoba, Ciudad Universitaria, 5000 C\'{o}rdoba, Argentina.}
\author{H. M. Pastawski}
\email{horacio@famaf.unc.edu.ar}
\affiliation{Facultad de Matem\'{a}tica, Astronom\'{\i}a y F\'{\i}sica, Universidad
Nacional de C\'{o}rdoba, Ciudad Universitaria, 5000 C\'{o}rdoba, Argentina.}

\begin{abstract}
The evolution of a surface excitation in a two dimentional model is analyzed.
I) It starts quadratically up to a spreading time $t_{S}.$ II) It follows an
exponential behavior governed by a \textit{self-consistent Fermi Golden Rule}.
III) At longer times, the exponential is overrun by an inverse power law
describing return processes governed by quantum diffusion. At this last
transition time $t_{R}$ a\textit{\ survival collapse} becomes possible,
bringing the survival probability down by several orders of magnitude. We
identify this strongly destructive interference as an antiresonance in the
time domain.

\end{abstract}
\maketitle

\baselineskip=12pt

\section{Introduction}

We consider the dynamics of a charge excitation in a typical model for Tamm
states \cite{DS96}. Similar tight-binding models \cite{PM01} are used to
describe a variety of situations: molecules absorbed in metallic substrates,
decay of high-energy electron excitations and the decoherence caused by a weak
interaction with an \textquotedblleft environment\textquotedblright\ whose
spectrum is dense. The decay of the survival probability $P_{00}(t)$ of the
resulting resonant state is usually described, within a Markovian
approximation, by the Fermi Golden Rule (FGR). However, this description
contains approximations that leave aside some intrinsically quantum behaviors.
Various works on models for nuclei, composite particles \cite{Kha58, FGR78,
GMM95}, excited atoms either in a free electromagnetic field \cite{FP99} or in
photonic lattices \cite{KKS94}, showed that the exponential decay has
superimposed beats and does not hold for very short and very long times,
compared with the lifetime of the system.

In Ref. \cite{RP05} we presented a model describing the evolution of a surface
excitation in a semi-infinite chain, a model that is solved analytically and
susceptible for an experimental test \cite{MBS+97}. Here, we present a general
analysis showing the quantum nature of the deviations from the Fermi Golden
Rule. Then, we numerically\emph{\ }solve a model consisting of an excited add
atom in a two dimensional lattice. We identify three time regimes in the decay
of the survival probability $P_{00}(t)$:\ 1) For short times the decay is
quadratic, as is expected when the coupling of the local state with the
continuum is perturbative. 2) An intermediate regime characterized by an
exponential behavior, the \textit{self-consistent Fermi Golden Rule} (SC-FGR)
where the rate, the pre-exponential factor and the characteristic
frecuency\ are found self-consistently. 3) A long-time regime in which the
exponential decay of the\textit{\ pure survival} probability is overrun by an
inverse power law, which is identified with the \textit{return} probability
enabled by the slow quantum diffusion in the substrate. At this last
cross-over, the oscillations could lead to a dip in $P_{00}(t)$ of several
orders of magnitude. This \textit{survival collapse} is identified with a
destructive interference between the \textit{pure survival} amplitude, i.e.,
the SC-FGR component, and the \textit{return} amplitude, associated with high
orders in a perturbation theory.

\section{Survival probability of a surface excitation}

We consider the evolution of a surface excitation, prepared in the state
$\left\vert 0\right\rangle $ in a Hamiltonian with finite spectrum as is the
case of most excitations in a lattice. The survival probability is
\begin{align}
P_{00}\left(  t\right)   &  =\left\vert \left\langle 0\right\vert
\exp[-\mathrm{i}\hat{H}t/\hbar]\left\vert 0\right\rangle \theta\left(
t\right)  \right\vert ^{2}\\
&  \equiv\hbar^{2}\left\vert G_{00}^{R}\left(  t\right)  \right\vert ^{2},
\end{align}
where
\begin{equation}
G_{00}^{R}\left(  t\right)  =\int\frac{\mathrm{d}\varepsilon}{2\pi\hbar}%
G_{00}^{R}(\varepsilon)\exp[-\mathrm{i}\varepsilon t/\hbar], \label{Eq.GtGe}%
\end{equation}
is the retarded Green's function for a single fermion. Expanding the initial
condition in the eigenstates of $\hat{H}$ yields \cite{KF47, Kha58}
\begin{equation}
P_{00}(t)=\left\vert \theta\left(  t\right)  \int_{-\infty}^{\infty}%
\mathrm{d}\varepsilon\text{ }N_{0}\left(  \varepsilon\right)  \exp
[-\mathrm{i}\varepsilon t/\hbar]\right\vert ^{2}, \label{Eq_Pcomun}%
\end{equation}
where $N_{0}\left(  \varepsilon\right)  $ is the Local Density of States
(LDoS) at site $0$th which, in terms of the retarded Green%
\'{}%
s function is $N_{0}\left(  \varepsilon\right)  =-1/\pi\operatorname{Im}%
G_{00}^{R}(\varepsilon).$ Hence, we can\textbf{\ }evaluate the survival
probability using the Fourier transform of the LDoS, which can be accurately
calculated in the energy representation where the integral is limited to the
spectral support. Besides, a clear identification of quantum interferences
will be obtained by analyzing the argument of the square modulus.%
\begin{figure}
[ptb]
\begin{center}
\includegraphics[
trim=0.458482in 0.671674in 0.810695in 0.178034in,
height=2.1119in,
width=3.1773in
]%
{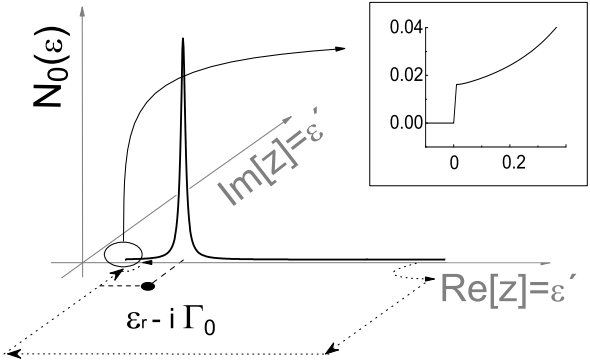}%
\caption{Local spectrum (LDoS) in the complex plane $z=\varepsilon+$i
$\varepsilon\acute{}$. $\varepsilon_{L}$ and $\varepsilon_{U}$ are the lower
and upper band-edges, respectively. The pole appears in $\varepsilon
_{r}-\mathrm{i}\Gamma_{0}.$ The integration path is shown with dotted lines;
consist of four straight lines and two arcs, that avoid the band-edges
singularities.}%
\label{fig_camino2}%
\end{center}
\end{figure}

In order to evaluate the local dynamics, we perform the integral in
Eq.(\ref{Eq_Pcomun}) using the residue theorem and following the path shown in
the Fig. \ref{fig_camino2}. In the analytical continuation $N_{0}(z)\equiv
N_{0}(\varepsilon+\mathrm{i}\varepsilon^{\prime})$, resonances appear as poles
in the complex plane. We will consider Hamiltonians where an initially
unperturbed state of energy $\varepsilon_{0}=\left\langle 0|\hat
{H}|0\right\rangle $ interacting with a continuum is a well defined resonance,
i.e., the expansion of $\left\vert 0\right\rangle $ in terms of the
eigenstates has a small breath $\Gamma_{0}$ around an energy $\varepsilon
_{r}=\varepsilon_{0}+\Delta_{0}$, where $\Delta_{0}=\Delta\left(
\varepsilon=\varepsilon_{r}\right)  $ is a small shift due to the interaction.
We exclude systems with localized eigenstates \cite{Ander}. Then%
\begin{align}
P_{00}(t)  &  =|\underset{\mathrm{SC-FGR}}{\underbrace{~a~~\mathrm{e}%
^{-(\Gamma_{0}+\mathrm{i}\varepsilon_{r})t/\hbar}}}\nonumber\\
&  +\int\limits_{0}^{\infty}\left[  \mathrm{e}^{-\left(  \varepsilon^{\prime
}+\mathrm{i}\text{ }\varepsilon_{L}\right)  t/\hbar}N_{0}(\varepsilon
_{L}-\mathrm{i}\varepsilon^{\prime})\right. \nonumber\\
&  \underset{\text{return correction from quantum diffusion}}{\underbrace
{\left.  -\mathrm{e}^{-\left(  \varepsilon^{\prime}+\mathrm{i}\text{
}\varepsilon_{U}\right)  t/\hbar}N_{0}(\varepsilon_{U}-\mathrm{i}%
\varepsilon^{\prime})\right]  \text{\textrm{d}}\varepsilon^{\prime}}}|^{2},
\label{Eq_Poo}%
\end{align}
where
\begin{align}
a  &  =2\pi\mathrm{i}\lim_{z\rightarrow\varepsilon_{r}-\mathrm{i}\Gamma_{0}%
}\left[  (z-\varepsilon_{r}+\mathrm{i}\Gamma_{0})\text{ }N_{0}(z)\right]  ,\\
&  =2\pi\mathrm{i}\left[  1-\left.  \frac{\partial}{\partial\varepsilon}%
\Delta\left(  \varepsilon\right)  \right\vert _{\varepsilon_{r}-\mathrm{i}%
\Gamma_{0}}\right]  ^{-1}.
\end{align}
The first term of Eq.(\ref{Eq_Poo}) already supersedes the usual Fermi Golden
Rule approximation since it has a pre-exponential factor ($|a|^{2}\gtrsim1$)
and the exact rate of decay $\Gamma_{0}$. This result is the
\textit{self-consistent Fermi Golden Rule} (SC-FGR). By analogy with a
classical Markov model, this exponential term is identified with a
\textit{\textquotedblleft pure survival\textquotedblright\ amplitude. }Within
the same analogy\textit{,} the second term will be
called\textit{\ \textquotedblleft return\textquotedblright\ amplitude, }as it
is fed upon the initial decay. The first term is the dominant one for a wide
range of times, while the \textquotedblleft quantum
diffusion\textquotedblright\ described by the second, dominates for long times
and brings out the details of the spectral structure of the system. The second
term is also fundamental for the normalization at very short times where the
most excited energy states of the whole system can be virtually explored. Both
terms combine to provide the initial quadratic decay required by the
perturbation theory:%
\begin{equation}
P_{00}\left(  t\right)  =1-\frac{t^{2}}{\hbar^{2}}\left\langle (\varepsilon
-\varepsilon_{r})^{2}\right\rangle _{N_{0}}+\cdots. \label{Eq_corto}%
\end{equation}
Here $\left\langle (\varepsilon-\varepsilon_{r})^{2}\right\rangle _{N_{0}}$ is
the energy second moment of the density $N_{0}\left(  \varepsilon\right)  $.
This expansion holds for a time shorter than the spreading time $t_{S}$ of the
wave packet formed upon decay.

For long times, the behavior of $P_{00}\left(  t\right)  $ is governed by the
slowly decaying second term in Eq.(\ref{Eq_Poo}). Only small values of
$\varepsilon^{\prime}$ contribute to the integral. This restricts the
integration of the LDoS to a range near the band-edges. Then, one can go back
to Eq.(\ref{Eq_Pcomun}) and perform the Fourier transform retaining only the
van Hove singularities \cite{SS75, FTW76} at these edges. The relative
participation of the energy states at each edge of the LDoS is given by the
relative weight of the Lorentzian tails at these edges $\beta=[(\varepsilon
_{r}-\varepsilon_{L})^{2}+\Gamma_{0}^{2}]/[(\varepsilon_{U}-\varepsilon
_{r})^{2}+\Gamma_{0}^{2}].$ Then, the survival probability for long times is
\begin{align}
P_{00}(t)  &  \approx\left[  1+\beta^{2}-2\beta\cos(Bt/\hbar)\right]
\nonumber\\
&  \times\left\vert \int\text{\textrm{d}}\varepsilon^{\prime}\mathrm{e}%
^{-\varepsilon^{\prime}t/\hbar}N_{0}(\varepsilon_{L}-\mathrm{i}\varepsilon
^{\prime})\right\vert ^{2}. \label{Eq_Plargo}%
\end{align}
This means that the long time behavior is just the power law decay of the
integral multiplied by a factor containing a modulation with frequency
$B/\hbar.$

\section{Survival collapse}

In steady state transport \cite{DPW89} as well as in dynamical electron
transfer \cite{LPD90} there are situations when a particle can reach the final
state following two alternative pathways. Since each of them collects a
different phase, this allows a destructive interference blocking the final
state. This phenomenon has been dubbed \textit{antiresonance} \cite{DPW89,
LPD90}. It extends the Fano resonances describing the anomalous ionization
cross-section \cite{Fa61}. In the present case, the survival of the local
excitation also recognizes two alternative pathways: the
\textit{pure\ survival} amplitude, which is typically described by the Fermi
Golden Rule, and the pathways where the excitation has decayed, explored the
substrate, and then returns. These two alternatives can interfere. We rewrite
Eq.(\ref{Eq_Poo}) to emphasize that the survival probability $P_{00}\left(
t\right)  $ is the result of \textit{two} different contributions:
\begin{align}
P_{00}\left(  t\right)   &  =\left\vert \Psi_{S}+\Psi_{R}\right\vert
^{2},\label{Eq_Pasar}\\
&  =\left\vert \Psi_{S}\right\vert ^{2}+\left\vert \Psi_{R}\right\vert
^{2}+2\operatorname{Re}[\Psi_{S}^{\ast}\Psi_{R}],
\end{align}
where the phase in $\Psi_{R}$ arise from the exponentials with $\varepsilon
_{L}$ and $\varepsilon_{U}$ (the LDoS is real for any argument). Hence,
\begin{align}
\Psi_{S}\left(  t\right)   &  =\left\vert a\right\vert \mathrm{e}%
^{-\mathrm{i}\phi_{a}}\mathrm{e}^{-\Gamma_{0}t/\hbar}\mathrm{e}^{-\mathrm{i}%
\left(  \varepsilon_{r}-\varepsilon_{L}\right)  t/\hbar},\\
\Psi_{R}\left(  t\right)   &  =\left\vert \Psi_{R}\left(  t\right)
\right\vert \mathrm{e}^{\mathrm{i}\phi\left(  t\right)  };\\
\phi\left(  t\right)   &  =\arctan\left(  \frac{\beta\sin\left(
Bt/\hbar\right)  }{1-\beta\cos\left(  Bt/\hbar\right)  }\right)  .
\label{Eq_phaser}%
\end{align}
where Eq.(\ref{Eq_phaser}) results using the long time limit of
Eq.(\ref{Eq_Plargo}). While the interference term in $P_{00}\left(  t\right)
$ is present along the whole exponential regime, it becomes important when
both, the \textit{pure survival}\emph{\ }amplitude and the \textit{return}
contribution, are of the same order. This occurs at the cross-over time
$t_{R}$ between the exponential regime and the power law. The interference
term can produce\ a \textit{survival collapse}, i.e., a pronounced dip that
takes $P_{00}\left(  t\right)  $ close to zero (see Fig. \ref{fig_semi2}). In
order to obtain a full collapse, two simultaneous conditions are needed;
\begin{align}
\left\vert \Psi_{S}\left(  t_{R}\right)  \right\vert  &  =\left\vert \Psi
_{R}\left(  t_{R}\right)  \right\vert \text{ \ and \ }\\
\left(  \varepsilon_{r}-\varepsilon_{L}\right)  t_{R}/\hbar-\phi\left(
t_{R}\right)   &  =\left(  \pi-\phi_{a}\right)  +2\pi n, \label{Eq_fases}%
\end{align}
which are satisfied with a fair precision because the return amplitude has a
phase with a slow variation:
\begin{equation}
\left\vert \left(  \varepsilon_{r}-\varepsilon_{L}\right)  /\hbar\right\vert
\gg\Gamma_{0}/\hbar>2\pi/t_{R}\geq\left\vert \phi\left(  t_{R}\right)
\right\vert /t_{R},
\end{equation}
while, the pure survival term oscillates rapidly. When both amplitudes are of
the same order, the destructive interference will be noticeable.

\section{Decay in a 2d-system}%

\begin{figure}
[ptb]
\begin{center}
\includegraphics[
height=1.0603in,
width=2.5183in
]%
{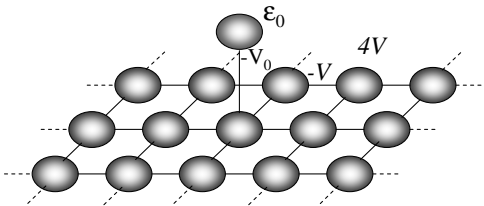}%
\caption{2d-lattice with an add atom with site energy $\varepsilon_{0}$ and
hopping $V_{0}.$}%
\label{Fig_esquema}%
\end{center}
\end{figure}

The above results (Eq.(\ref{Eq_Poo}), Eq.(\ref{Eq_Plargo}), and the survival
collapse effect) were verified and quantified in a recent publication
\cite{RP05}, which solves the dynamics of a surface spin excitation weakly
coupled to a semi-infinite chain of interacting spins. The integral in
Eq.(\ref{Eq_Pcomun}) is analytically solved for the different time regimes
(short, exponential and long time). Also, the cross-over from the short time
regime to the exponential SC-FGR $t_{S}$, and the cross-over from the SC-FGR
to the power law regime $t_{R}$, were found. The survival collapse takes place
at $t_{R}$.

Here, we shall consider a square lattice with an add atom. The Tight Binding
Hamiltonian is%
\begin{equation}
\hat{H}=%
{\displaystyle\sum\limits_{n}}
\left\vert n\right\rangle \varepsilon_{n}\left\langle n\right\vert -%
{\displaystyle\sum\limits_{n,m}}
\left\vert n\right\rangle V_{n,m}\left\langle m\right\vert ,
\end{equation}
where each $\left\langle r|n\right\rangle $\ is centered around the
corresponding lattice site $n$, $\varepsilon_{n}$ are the site energies and
$V_{n,m}$ are the hoppings. We consider the case where the $0$th site denotes
the add atom, i.e., is different from the others sites in both site energy,
$\varepsilon_{0}\neq\varepsilon_{n}\equiv4V$ and hopping $V_{0,1}\equiv
V_{0}<V_{n,m}\equiv V$. This defines a continuous spectrum in the range
$[0,B\equiv8V]$\textbf{.}\ The Green function of this problem, and hence the
LDoS, is evaluated using the Dyson equation
\begin{equation}
\left[  G_{00}^{R}\left(  \varepsilon\right)  \right]  ^{-1}=\left[
G_{00}^{R\left(  0\right)  }\left(  \varepsilon\right)  \right]  ^{-1}%
+V_{0,1}G_{11}^{R\left(  0\right)  }\left(  \varepsilon\right)  V_{1,0},
\end{equation}
following the general continued fraction procedure described in Ref.
\cite{PM01}:
\begin{equation}
G_{00}^{R}(\varepsilon)=\frac{1}{\varepsilon-\varepsilon_{0}-V_{0}^{2}%
G_{11}^{R\left(  0\right)  }(\varepsilon)}, \label{Eq_Goo2d}%
\end{equation}
where $G_{11}^{R\left(  0\right)  }(\varepsilon)$ is the Green function for a
periodic square lattice (Ref. \cite{Eco79}).

For this system, the local second moment of the Hamiltonian is $V_{0}^{2}.$
The short time regime (Eq.(\ref{Eq_corto})) holds up to a time $t_{S}$ in
which the quadratic decay becomes an exponential. A good estimate of $t_{S}%
$\ is obtained from the minimal distance between the short time decay and the
exponential decay. We can use the first pole approximation (evaluating
$G_{11}^{R\left(  0\right)  }\left(  \varepsilon=\varepsilon_{0}\right)  $ in
Eq.(\ref{Eq_Goo2d})) to obtain $\Gamma_{0}\approx\pi V_{0}^{2}N_{1}^{\left(
0\right)  }\left(  \varepsilon_{0}\right)  ,$\ which coincides with the FGR.
In the same order of approximation we can take $|a|^{2}\approx1$
in\ Eq.(\ref{Eq_Poo}). Then we obtain\emph{\ }%
\begin{equation}
t_{S}\approx\hbar\pi N_{1}^{\left(  0\right)  }\left(  \varepsilon_{0}\right)
.
\end{equation}
This result shows that the \textit{spreading} time $t_{S}$ is only determined
by $N_{1}^{\left(  0\right)  }\left(  \varepsilon_{0}\right)  ,$\ the local
density of states at the first site of the\ unperturbed substrate.

\begin{figure}
[ptb]
\begin{center}
\includegraphics[
trim=0.540658in 0.308312in 0.303548in 0.362340in,
height=2.8236in,
width=3.5803in
]%
{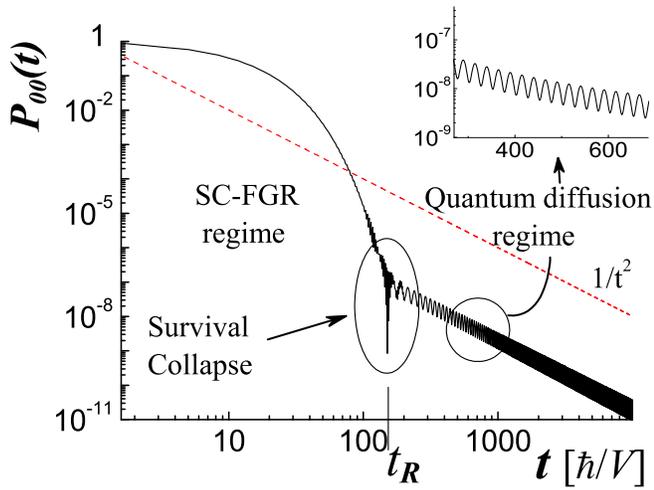}%
\caption{Local polarization, in a doble logarithmic scale, as a function of
time. We consider an unperturbed energy of $\varepsilon_{0}/V=2$ and
interaction strength $V_{0}/V=0.4.$ This is the case that we consider in
Fig.\ref{fig_camino2}. The decay exhibits: The exponential behavior as
described by the self-consistent Fermi Golden Rule, and an asymptotic square
power law decay. The inset shows the oscillation that modulates this decay.
The cross-over time $t_{R}$ when the survival collapse takes place is
indicated.}%
\label{fig_semi2}%
\end{center}
\end{figure}

We verify Eq.(\ref{Eq_Poo}), Eq.(\ref{Eq_Plargo}), and the survival collapse
effect, using the analytical expression for $G_{00}^{R}\left(  \varepsilon
\right)  \ $and $N_{0}(\varepsilon),$ and performing the numerical Fourier
transform. In Fig. \ref{fig_semi2} we show $P_{00}(t)$. The curve shows the
exponential SC-FGR, which then is overrun by a $1/t^{2}$\ power law decay. The
cross-over time $t_{R}$ is easily identified through the survival collapse
shown as a dip in $P_{00}$. There, the survival probability suddenly decreases
from its average by almost three orders of magnitude. The inset shows the
small oscillation that modulates the power law.

It is important to note that the square power law decay obtained for long
times is a consequence of the $\theta\left(  \varepsilon\right)  $ dependence
of the LDoS (inset of Fig. \ref{fig_camino2}), i.e., this power law\ is
consistent with Eq.(\ref{Eq_Plargo}) taken together with Eq.(\ref{Eq_Goo2d}).%

\section{Conclusions}

In the present work we have discussed the dynamics of a local excitation that
decays through a weak interaction with a continuum spectrum with finite
support. Our approach goes beyond the usual Markovian approximation that uses
the Fermi Golden Rule to describe these environmental interactions.

The evolution starts with the expected quadratic decay. Then, it follows the
usual exponential FGR regime, but with a corrected rate and a pre-exponential
factor, i.e., the SC-FGR. Finally, we get the long time regime, that consists
of a square power law decay modulated by oscillations whose frequency is
determined by the bandwidth. This power law decay is a consequence of the
$\theta\left(  \varepsilon\right)  $ behavior of the LDoS in the band-edge
(Eq.(\ref{Eq_Plargo})) and is identified with the quantum diffusion in the
substrate. Hence, anomalies in the excitation decay gives information about
the substrate dynamics.

Finally, we predict the existence of the survival collapse. This non-Markovian
result fully considers the memory effects to infinite order. Such effect,
hinted but not explained in previous works, is visualized as the destructive
interference between the pure survival amplitude and the return amplitude that
arises from pathways that have already explored the rest of the system.

\end{document}